\documentclass[twocolumn, showpacs,  floatfix,aps,a4paper,prd]{revtex4}
\usepackage{graphicx}

\def \beq {\begin{equation}}
\def \eeq {\end{equation}}
\def \beqn {\begin{eqnarray}}
\def \eeqn {\end{eqnarray}}

\begin{document}

\title{Local electromagnetic duality and gauge invariance}

\author{Alberto Saa}
\email{asaa@ime.unicamp.br}
\affiliation{
Departamento de Matem\'atica Aplicada,
UNICAMP,
  Campinas, SP, Brazil}

\begin{abstract}
Bunster and Henneaux and, separately, Deser have very recently considered the possibility of gauging the usual   electromagnetic duality of Maxwell equations. By using   off-shell manipulations  in the context of the Principle of least action, they conclude that this  is not  possible, at least with the conventional compensating gauge fields scheme.
Such a conclusion contradicts, however, an old result of Malik and Pradhan, who showed that it is indeed possible to introduce an extra   abelian gauge   field directly in the vacuum Maxwell equations in order to render them covariant under local duality transformations.
Since it is well known that the equations of motion can, in general, admit
more symmetries than the associate Lagrangian, this would not be
 a paradoxal result,
in principle.
Here, we revisit these works and identify the source of the
different conclusions. We show that the Malik-Pradhan procedure does not
preserve the original Maxwell gauge invariance, while Bunster, Henneaux, and Deser  sought for generalizations which are, by construction, invariant under the Maxwell gauge transformation. Further,
we show  that the Malik-Pradhan construction cannot be adapted or extended
 in order
to preserve the Maxwell gauge invariance, reinforcing  the conclusion that it is not possible to gauge the
electromagnetic duality.
\end{abstract}


\maketitle

\section{Introduction}
It is well known that Maxwell equations in vacuum (natural units are adopted through this work)
\beqn
\label{Maxwell}
\nabla\cdot \mathbf{E} &=& 0, \quad \frac{\partial}{\partial t} \mathbf{B} + \nabla\times \mathbf{E} = 0,  \\
\nabla\cdot \mathbf{B} &=& 0,\quad \frac{\partial}{\partial t} \mathbf{E} - \nabla\times \mathbf{B} =0,\nonumber
\eeqn
are invariant under the electromagnetic duality transformation \cite{Jackson,Olive:1995sw}
\beqn
\label{duality}
\mathbf{E} &\to& \mathbf{E}' = \cos\theta \mathbf{E} - \sin\theta \mathbf{B},  \\
\mathbf{B} &\to&  \mathbf{B}' = \cos\theta \mathbf{B} + \sin\theta \mathbf{E}\nonumber .
\eeqn
The study of the electromagnetic duality has led to numerous advances in quantum mechanics and field theory, ranging from the   Dirac charge quantization   to the introduction of S-duality in string-inspired models, see, for instance, \cite{Olive:1995sw,AlvarezGaume:1996mv}. Very recently, Bunster and Henneaux \cite{Bunster:2010wv}
and, separately, Deser \cite{Deser:2010it} investigated the problem of gauging the electromagnetic duality, {\em i.e.}, to consider that, instead of the rigid rotation (\ref{duality}),   the angle $\theta$ could depend on the spacetime position. By using off-shell manipulations in the electromagnetic action as done, for instance, in \cite{Deser:1976iy,Deser:1982}, they conclude that the electromagnetic duality cannot be gauged, at least by means of the conventional compensating gauge fields scheme.

Nevertheless, exactly the same question about the gauging of the duality (\ref{duality}) was raised 25 years ago by Malik and Pradhan \cite{Malik:1984cr}. Surprisingly, they show that the Maxwell equations (\ref{Maxwell}) can indeed be modified by introducing a compensation massless vector field  in order to accommodate a local duality transformation. Their analysis was based on the equations of motion, and not on the Principle of least action as the more recent works \cite{Bunster:2010wv,Deser:2010it} were. Since the equations of motion can have in general more symmetries than the corresponding Lagrangian \cite{Henneaux:1984ke,Morandi:1990su,Olver}, such a result would not be, in principle, a paradox.
Let us briefly recall the Malik-Pradhan construction in order to clarify why they got   a distinct conclusion. For our purposes here, it is more convenient  to introduce the complex electromagnetic vetor
\beq
\label{complex}
\mathbf{F} =  \mathbf{E} + i \mathbf{B},
\eeq
which, incidentally, can be traced back to the XIX century, see\cite{Olive:1995sw,laporte}. Maxwell equations in vacuum for the complex electromagnetic vetor (\ref{complex}) read simply as
\beq
\label{ComplexMaxwell}
\nabla\cdot \mathbf{F} = 0, \quad
 i\frac{\partial}{\partial t} \mathbf{F} - \nabla\times \mathbf{F} = 0,
\eeq
and the local duality transformation will be given by
\beq
\label{localF}
 \mathbf{F}   \to \mathbf{F}' =  e^{i\theta(x)}\mathbf{F}.
\eeq
It is clear that (\ref{ComplexMaxwell}) is not invariant nor   covariant under the local duality
transformation (\ref{localF}) since
\beqn
  {\partial_t} \mathbf{F} \to   {\partial_t} \mathbf{F}' &=& e^{i\theta(x)}\left(  {\partial_t} \mathbf{F} + i {\partial_t}\theta\  \mathbf{F}\right),
  \nonumber \\
\nabla\cdot \mathbf{F} \to \nabla\cdot \mathbf{F}' &=& e^{i\theta(x)}\left( \nabla\cdot \mathbf{F} + i\nabla\theta\cdot \mathbf{F}\right), \\
\nabla\times \mathbf{F} \to \nabla\times \mathbf{F}' &=& e^{i\theta(x)}\left( \nabla\times \mathbf{F} + i\nabla\theta\times \mathbf{F}\right).  \nonumber
\eeqn
However, this obstruction can be easily circumvented by introducing a real four vector $a_\mu = (a_0, \mathbf{a})$ transforming as
\beq
\label{gauge}
a_\mu \to a'_\mu = a_\mu + \partial_\mu \theta
\eeq
under (\ref{localF}). The operators ${\cal D}_t = \partial_t - ia_0$ and
${\cal D}  = \nabla - i\mathbf{a}$ transform under (\ref{localF}) and (\ref{gauge}) in the correct way
\beqn
  {{\cal D}_t} \mathbf{F} \to   {{\cal D}'_t} \mathbf{F}' &=& e^{i\theta(x)}   {{\cal D}_t} \mathbf{F}  ,
  \nonumber \\
{\cal D}\cdot \mathbf{F} \to {\cal D}'\cdot \mathbf{F}' &=& e^{i\theta(x)} {\cal D}\cdot \mathbf{F},    \\
{\cal D}\times \mathbf{F} \to {\cal D}'\times \mathbf{F}' &=& e^{i\theta(x)}  {\cal D}\times \mathbf{F}  , \nonumber
\eeqn
rendering the equations
\beqn
\label{MalikPradhan1}
\left(\nabla - i\mathbf{a} \right)\cdot \mathbf{F} &=& 0,  \\
i\left(  \frac{\partial}{\partial t} - ia_0\right) \mathbf{F} - \left(\nabla - i\mathbf{a} \right)\times \mathbf{F} &=& 0,
\label{MalikPradhan2}
\eeqn
covariant under the transformations (\ref{localF}) and (\ref{gauge}). Eqs. (\ref{MalikPradhan1}) and  (\ref{MalikPradhan2}) have interesting properties which we shall discuss  later, but they have a serious drawback lying at the heart of the contradiction between the works \cite{Bunster:2010wv,Deser:2010it} and \cite{Malik:1984cr}: the usual Maxwell gauge invariance is irremediably lost.

\section{Two gauge invariances}

The covariance of Malik-Pradhan (MP) generalized Maxwell equations (\ref{MalikPradhan1}) and  (\ref{MalikPradhan2}) under the local
 duality transformation (\ref{localF}) and (\ref{gauge}) can be summarized as follows: if
$(a_\mu, \mathbf{F})$ is a valid solution of the MP equations, then
$(a_\mu + \partial_\mu\theta, e^{i\theta}\mathbf{F})$ is another one. The four-vector field $a_\mu$ behaves as an abelian gauge field and, hence, any quantity based on the tensor
\beq
\label{f}
f_{\mu\nu} = \partial_\mu a_\nu - \partial_\nu a_\mu
\eeq
is invariant under the local duality gauge invariance.
The dynamics for the gauge field $a_\mu$ can be introduced
by exploring the invariant tensor (\ref{f}).
There are two evident electromagnetic invariants under the local duality
transformation, namely the electromagnetic energy density and the Poynting vector
\beqn
\frac{1}{2} {\mathbf{F}}^*\cdot \mathbf{F} &=& \frac{1}{2}\left( |\mathbf{E}|^2 + |\mathbf{B}|^2\right),\\
 \frac{1}{2i}{\mathbf{F}}^*\times \mathbf{F}&=&   \mathbf{E}\times \mathbf{B} ,\nonumber
\eeqn
where   $ {\mathbf{F}}^* = \mathbf{E} - i\mathbf{B}$.
Several other interesting properties of (\ref{MalikPradhan1}) and (\ref{MalikPradhan2}) are discussed in \cite{Malik:1984cr}.

The MP equations
(\ref{MalikPradhan1}) and (\ref{MalikPradhan2}) are a legitimate generalization
of Maxwell equation with respect to the local electromagnetic duality. Moreover, they can be considered as the simplest possible generalization,
obtained by means of the usual
    minimal coupling procedure, with the introduction of the covariant derivative $\partial_\mu\to \partial_\mu - ia_\mu $.
 However,
they lost their distinctive Maxwell gauge invariance. A solution of the usual Maxwell equations corresponds to a $(0,\mathbf{F})$ solution    of
the MP equations. For $a_\mu =0$, Eq. (\ref{MalikPradhan1}) implies that
\beq
\label{vectorpot}
\mathbf{F} = \nabla\times \mathbf{A},
\eeq
where $\mathbf{A}$ is a complex  vector potential with real and complex parts corresponding, respectively, to the electric and magnetic vector potentials. The
electromagnetic vector defined by (\ref{vectorpot}) is invariant under the
usual Maxwell gauge transformation
\beq
\label{Maxgauge}
\mathbf{A} \to \mathbf{A}' = \mathbf{A} + \nabla\phi,
\eeq
for a complex function $\phi$. In the variational formulation considered in
\cite{Bunster:2010wv,Deser:2010it}, the
first complex Maxwell equation in (\ref{ComplexMaxwell}) and its
solution (\ref{vectorpot})
are
  taken for granted, and   the second complex Maxwell equation is then obtained
  from the minimization of the electromagnetic action with respect to
  the variable $\mathbf{A}$, assuring by construction, in this way, the invariance of the whole system under (\ref{Maxgauge}). It is  clear now the origin of the
  distinct conclusions: while Bunster and Henneaux \cite{Bunster:2010wv}, and Deser \cite{Deser:2010it} have been looking for
  extensions of the gauge-invariant electromagnetic action, Malik and Pradhan
  \cite{Malik:1984cr} modified all the Maxwell equations, but spoiled the
  Maxwell gauge invariance.

The Malik-Pradhan construction is, indeed, incompatible with the Maxwell gauge
invariance (\ref{Maxgauge}). First, Eq. (\ref{MalikPradhan1}) does not
admit (\ref{vectorpot}) as a solution in general and, consequently, the transformation (\ref{Maxgauge}) cannot be a symmetry of the system. One could try to modify  the MP construction by adopting a procedure
similar to that one used in the Principle
 of least action, attempting   to keep
the Maxwell gauge invariance by the construction. Let us assume
the form (\ref{vectorpot}) and look for a generalization of the second complex
Maxwell equation. If the electromagnetic vector transform as $\mathbf{F}\to
e^{i\theta}\mathbf{F}$ under the duality transformation
\beq
\label{MPgauge}
 \mathbf{A} \to \mathbf{A}'= e^{i\theta(x)}\mathbf{A},
\eeq
Eq. (\ref{MalikPradhan2}) will be the right generalization of the second
Maxwell equation. This can be achieved by   generalizing   (\ref{vectorpot})
to
  \beq
\label{vectorpot1}
\mathbf{F} = \left(\nabla -i\mathbf{a}\right)\times \mathbf{A},
\eeq
and taking into account (\ref{gauge}). However, (\ref{vectorpot1}) is not invariant under (\ref{Maxgauge}) anymore
and, moreover, it cannot be cast in covariant
form by introducing gauge compensation fields guided by the minimal coupling
procedure.
Under (\ref{Maxgauge}), we have $\mathbf{F}\to \mathbf{F} -i\mathbf{a}\times
\nabla\phi$ from (\ref{vectorpot1}). Such an extra term could be absorbed with the introduction of a
gauge scalar field, but in this case the local duality invariance would be
spoiled. In fact, we can show that
no local theory of this type can be simultaneously invariant under the
gauge transformations $\mathbf{A} \to  \mathbf{A} + \nabla\phi$
and covariant under $\mathbf{A} \to   e^{i\theta}\mathbf{A}$. Suppose we have two
equivalent potencials with respect to (\ref{Maxgauge}), say $\mathbf{A}$ and
$\mathbf{A} + \nabla\phi$. Now, applying for these two equivalent potentials
the gauge transformation (\ref{MPgauge}) we end up, respectively, with
$e^{i\theta}\mathbf{A}$ and $e^{i\theta}\mathbf{A} + e^{i\theta}\nabla\phi$,
which do not belong anymore to the same equivalent class since
$e^{i\theta}\nabla\phi \ne \nabla\phi'$ in general.
Notice that,
if we give up the generalization (\ref{vectorpot1}),
we will have no transformation
$\mathbf{A}\to
 \mathbf{A}'$ implying
$\mathbf{F}\to
e^{i\theta}\mathbf{F}$ since it would suppose that
\beq
\nabla\times \mathbf{A}' = e^{i\theta(x)}\nabla\times \mathbf{A},
\eeq
which has no solution for general $\theta(x)$ either. It is not possible
to keep the Maxwell gauge invariance along the  Malik-Pradhan construction,
 reinforcing Bunster, Henneaux, and Deser recent conclusion that it is not possible to gauge the
electromagnetic duality.

It is interesting to
notice  that, giving up of a local theory, it is indeed possible \cite{elcio} to construct a formal
solution of (\ref{MalikPradhan1}) simultaneously covariant under (\ref{localF}) and invariant under (\ref{Maxgauge})
  \beq
\label{elcio}
\mathbf{F}_{\Gamma_x} = \exp\left(i\int_{\Gamma_x} a_\mu d x^\mu\right)\nabla\times \mathbf{A},
\eeq
with the assumption that $\mathbf{A}$ is invariant under the local duality,
where $\Gamma_x$ is a continuous world line ending at the spacetime point $x=(t,\mathbf{x})$. However, for general $a_\mu$, the electromagnetic
vector (\ref{elcio}) does depend on the integration path ${\Gamma_x}$. Such
non-locality implies
severe non-uniqueness, challenging  the physical applicability of non-local
extensions of this type. On the other hand, for the case of pure gauge field $a_\mu=\partial_\mu\theta$, Eq. (\ref{elcio}) defines a unique electromagnetic
vector $\mathbf{F}  = e^{i\theta}\nabla\times \mathbf{A}$.

\section{Final remarks}

Even though the serious drawback of the breaking of  Maxwell gauge invariance, the
Malik-Pradhan equations (\ref{MalikPradhan1}) and (\ref{MalikPradhan2}) have
some interesting properties. In terms of the original electric and magnetic
fields, they read
\beqn
\label{MP3}
\nabla\cdot\mathbf{E} = - \mathbf{a}\cdot \mathbf{B}  , \quad   \nabla\cdot\mathbf{B} =\mathbf{a}\cdot \mathbf{E} \nonumber   \\
\frac{\partial}{\partial t} \mathbf{B}
+ \nabla\times \mathbf{E}  = a_0\mathbf{E}- \mathbf{a}\times \mathbf{B}, \\
\frac{\partial}{\partial t} \mathbf{E}
- \nabla\times \mathbf{B}  = -a_0\mathbf{B} + \mathbf{a}\times \mathbf{E}. \nonumber
\eeqn
Without loss of generality,
one can set $a_0=0$ by exploring the local duality transformations.
Malik and Pradhan suggested, in this case, a connection between (\ref{MP3})
and the Maxwell equations in a magneto-electric medium, {\em i.e.},
a material medium with a linear and reciprocal  relationship between the magnetic field and the electric polarization, and between the electric field and the magnetic polarization, besides the usual relationships between the magnetic field and the magnetic polarization and between the electric field and electric polarization\cite{magneto}
\beqn
\label{medium}
\mathbf{H} &=& \mu^{-1}\mathbf{B}  - \beta(\mathbf{x}) \mathbf{E},  \\
\mathbf{D} &=& \epsilon\mathbf{E}  + \beta(\mathbf{x}) \mathbf{B}. \nonumber
\eeqn
However, Eq. (23) in \cite{Malik:1984cr} requires, necessarily, the existence
of magnetic charges and currents, compromising the physical interpretation
of (\ref{MP3}) in terms of realistic electrodynamics in material media. In order
to save such an interpretation, we can restrict our analysis to solutions
such that $\mathbf{E}$ and $\mathbf{B}$ are orthogonal and
    $\mathbf{a}$ is parallel to $\mathbf{B}$. Compare now this
      particular case of
    (\ref{MP3}) with the Maxwell equations in a medium obeying
  (\ref{medium})
\beqn
\nabla\cdot \mathbf{B} &=& 0,   \nonumber \\
\epsilon \nabla\cdot \mathbf{E} &=& -\nabla\beta\cdot\mathbf{B}  \nonumber
,\\
\frac{\partial}{\partial t} \mathbf{E} - \nabla\times \mathbf{B} &=&0 , \\
  \mu^{-1} \frac{\partial}{\partial t} \mathbf{B} +
  \epsilon \nabla\times \mathbf{E} &=& -\nabla\beta\times\mathbf{B} .\nonumber
\eeqn
The MP equations  for such class of solutions are identical to the
Maxwell equations in a
  magneto-eletric medium such that $\mu^{-1}=\epsilon$ and
  $\mathbf{a}=\epsilon^{-1}\nabla\beta$. Notice that, as in the example considered
  by Malik and Pradhan, the vector $\mathbf{a}$ is a pure gauge field in this
  case, suggesting that the space dependent
  mixing of the electric and magnetic fields
  in (\ref{medium}) is, in fact,
  nothing more than
  a gauge effect! This point certainly deserves
  a deeper investigation, mainly due to the experimental relevance of
  magneto-electric media, see \cite{Fiebig} for a recente review.

Yet for the case of a pure gauge field $a_\mu=\partial_\mu\theta$, we have another interesting geometrical interpretation for the MP construction. The covariant derivatives with respect to the
 local duality can be rewritten, in this case,
by using the differential operator
\beq
{\cal D}_\mu = \partial_\mu -i\partial_\mu\theta = e^{i\theta}
\partial_\mu e^{-i\theta}.
\eeq
Such  kind of covariant derivatives are related to deformations of the
spacetime volume element and they have already been used to study the
compatibility between  the minimal coupling procedure and the Principle of least
action for
gauge fields in Riemann-Cartan spacetimes \cite{Saa}. In particular,
 the (complex) scalar field $\theta$ has a similar dynamics    to the
dilaton field\cite{geom}. Notice that another particular case
 of the MP construction involving
  a pure gauge field $a_\mu=\partial_\mu\theta$  was
independently proposed  by Kato and Singleton in \cite{Kato:2001zf}, where
they also
recognize the incompatibility between
  invariance under the usual
gauge transformations $\mathbf{A} \to  \mathbf{A} + \nabla\phi$
and covariance  under the local duality
 transformation $\mathbf{A} \to   e^{i\theta}\mathbf{A}$.

We finish  this note with two remarks. First, it is clear that the Malik-Pradhan construction is intrinsically  four (spacetime) dimensional due to the use of many
three (space) dimensional identities in its derivation. Indeed, one does not expect any electromagnetical
 duality transformation for other dimensions, only in four spacetime dimensions one can have
a duality between the magnetic and electric three-vectors. Second, in contrast to the original Maxwell equations,
in the Malik-Pradhan equations the symmetry between
self-dual ($\mathbf{F}^*=0$) and anti-self-dual ($\mathbf{F}=0$) solutions
is broken, {\em i.e.}, if $(a_\mu, \mathbf{F})$ is a self-dual solution of
(\ref{MalikPradhan1}) and (\ref{MalikPradhan2}), the corresponding
anti-self-dual solution will be $(-a_\mu, \mathbf{F}^*)$. They are equivalent
up to a local duality transformation only if $a_\mu$ is a pure gauge, {\em i.e.}, only if $f_{\mu\nu}=0$.

\acknowledgments
This work was supported by FAPESP and CNPq. It is a pleasure to thank E. Abdalla, S. Deser, and D. Singleton
for enlightening discussions and useful communications.

\end{document}